\documentclass[letter]{aa} 
\usepackage{graphicx,rotating}
\usepackage{txfonts}
\usepackage{soul}
\usepackage{natbib}
\usepackage{ulem}
\usepackage{gensymb}
\usepackage{subfig}
\usepackage{color}
\newlength{\dminus}
\newsavebox{\mybox}

\newcommand{\minnumb}{\settowidth{\dminus}{$-$}\kern-\dminus$-$}
\newcommand{\spcn}[1]{\settowidth{\dminus}{0}\kern#1\dminus}
\begin{document}
\title{The NH$_2$D hyperfine structure revealed by astrophysical observations\thanks{Based on observations carried out with the IRAM 30m Telescope. IRAM is supported by INSU/CNRS (France), MPG (Germany) and IGN (Spain).}}
\titlerunning{NH$_2$D hyperfine structure}
\authorrunning{Daniel et al.}
\author{
F.~Daniel\inst{\ref{inst0},\ref{inst1}} 
\and L. H. Coudert \inst{\ref{inst5}}
\and A. Punanova \inst{\ref{inst6}}
\and J. Harju \inst{\ref{inst6},\ref{inst8}}
\and A. Faure\inst{\ref{inst0},\ref{inst1}} 
\and E. Roueff \inst{\ref{inst3}}
\and O. Sipil\"a \inst{\ref{inst6}}
\and P. Caselli \inst{\ref{inst6}}
\and R. G\"usten \inst{\ref{inst7}}
\and A. Pon\inst{\ref{inst6},\ref{inst9}}
\and J. E. Pineda\inst{\ref{inst6}}
} 
\institute{
Univ. Grenoble Alpes, IPAG, F-38000 Grenoble, France  \label{inst0} 
\and
CNRS, IPAG, F-38000 Grenoble, France \label{inst1} 
\and
LISA, UMR 7583 CNRS-Universites Paris Est Cr\'eteil et Paris Diderot, 61 Avenue du G\'en\'eral de Gaulle, F-94010 Cr\'eteil, France \label{inst5} 
\and
Max Planck Institute for Extraterrestrial Physics (MPE), Giessenbachstr. 1, 85748 Garching, Germany \label{inst6}
\and
Department of Physics, P.O. Box 64, 00014 University of Helsinki, Finland  \label{inst8}
\and
LERMA, Observatoire de Paris, PSL Research University, CNRS, UMR8112, F-75014, Paris, France\label{inst3}
\and
Max Planck Institute for Radioastronomie, Auf dem H\"ugel 69, 53121 Bonn, Germany \label{inst7}
\and
Department of Physics and Astronomy, The University of Western Ontario, London, Canada, N6A 3K7\label{inst9}
}

\date{Received; accepted}


\abstract
{The 1$_{11}$-1$_{01}$ lines of ortho and para--NH$_2$D (o/p-NH$_2$D),
  respectively at 86 and 110 GHz, are commonly observed to provide
  constraints on the deuterium fractionation in the interstellar
  medium. In cold regions, the hyperfine structure due to the nitrogen
  ($^{14}$N) nucleus is resolved. To date, this splitting is the only
  one which is taken into account in the NH$_2$D column density
  estimates.}
{We investigate how the inclusion of the hyperfine splitting
caused by the deuterium (D) nucleus affects the analysis of the rotational lines of NH$_2$D.}
{We present 30m IRAM observations of the above mentioned lines, as
  well as APEX o/p-NH$_2$D observations of the 1$_{01}$-0$_{00}$ lines at 333 GHz. 
The hyperfine patterns of the observed lines
were calculated taking into account the splitting induced by the D
nucleus. The analysis then relies on line lists that either neglect or take into
  account the splitting induced by the D nucleus.}
{The hyperfine spectra are first analyzed with a line list that
  only includes the hyperfine splitting due to the $^{14}$N nucleus.
  We find inconsistencies between the line widths of the
  1$_{01}$-0$_{00}$ and 1$_{11}$-1$_{01}$ lines, the latter being
  larger by a factor of $\sim$1.6$\pm0.3$.
  Such a large difference is unexpected given the two
sets of lines are likely to originate from the same region. We next employ
a newly computed line list for
  the o/p-NH$_2$D transitions, where the hyperfine
  structure induced by both nitrogen and deuterium nuclei is
  included. With this new line list, the analysis of the previous
  spectra leads to linewidths which are compatible.}
{Neglecting the hyperfine structure owing to D leads to
overestimate the linewidths of the 
o/p-NH$_2$D lines at 3 mm. The error for a cold molecular core is
about 50\%. This error propagates directly to the column density
estimate.  It is therefore recommended to take into account the
hyperfine splittings caused by both the $^{14}$N and D nuclei in
any analysis relying on these lines.}


\maketitle
%

\section{Introduction}
The first firm identification of singly deuterated interstellar
ammonia, NH$_2$D, was reported by \cite{olberg1985} 
towards three molecular clouds at 86 and 110~GHz, the
frequencies of the $J_{K_a,K_c}$=1$_{11}$-1$_{01}$
 lines of o/p-NH$_2$D, respectively.
The identification was unambiguous thanks to the narrow linewidths which allowed to resolve the
hyperfine splitting due to the nitrogen nucleus. These two lines have 
since been detected in a variety of
cold astronomical sources and they are routinely employed to derive
the (D/H) ratio in ammonia \citep[see e.g.][]{roueff2005}. This ratio is a
sensitive tracer of the physical conditions and provides
strong constraints to deuterium fractionation models.

In the previous studies by \citet{coudert2006,coudert2009} and
in the current spectroscopic databases
(JPL\footnote{spec.jpl.nasa.gov} and
CDMS\footnote{http://www.astro.uni-koeln.de/cdms/}), the line lists of
the NH$_2$D microwave spectra are based on the measurements by
\cite{delucia1975}, \cite{cohen1982} and \citet{fusina_88}. These data sets only consider
the nitrogen quadrupole hyperfine structure. In this letter, a new
theoretical analysis of the NH$_2$D hyperfine structure is presented
which takes into account the nitrogen quadrupole interaction, the
deuteron quadrupole interaction and the nitrogen spin-rotation
interaction. It is based on the measurements by \cite{kukolich1968}, \cite{cohen1982},
and \citet{fusina_88}. We show that the
inclusion of the three coupling terms in the analysis provides a
simple explanation for the origin of different linewidths in the
1$_{01}$-0$_{00}$ and 1$_{11}$-1$_{01}$ lines of NH$_2$D, as
observed recently towards the prestellar core H-MM1. 
Historically, the first astronomical observation resolving the hyperfine splitting owing
to the D quadrupole coupling was presented by \citet{caselli2005} for DCO$^+$.

The paper is organized as follows. In Sect. \ref{sec:observations}, we describe
the IRAM and APEX observations of the o/p--NH$_2$D lines detected towards H-MM1.
Sect. \ref{sec:spectro} gives the details of the spectroscopy calculations.
In Sect. \ref{sec:models}, we discuss the impact of the new line list on the derivation
of radiative transition parameters and we conclude in Sect. \ref{sec:conclusion}.

\section{Observations} \label{sec:observations}

The target of the present NH$_2$D observations is the dense, starless
core H-MM1 lying in the eastern part of Lynds 1688 in Ophiuchus
(\citealt{2004ApJ...611L..45J}; \citealt{2011A&A...528C...2P}). The
position observed, $\alpha$=$16^{\rm h}27^{\rm m}59\fs0$, $\delta$=$-24\degr33\arcmin33\arcsec$,
 was chosen as the H$_2$ column density peak, as derived from pipeline-reduced Herschel
far-infrared maps. These were downloaded from the Herschel Science
Archive\footnote{www.cosmos.esa.int/web/herschel/science-archive}. 

\subsection{APEX 12m}

The ground-state $1_{01}$--$0_{00}$ transitions of o/p-NH$_2$D
at $\sim$333 GHz were observed using the upgraded
version of the First Light APEX\footnote{This publication is based on data acquired with the Atacama Pathfinder Experiment (APEX). APEX is a collaboration between the Max-Planck-Institut fur Radioastronomie, the European Southern Observatory, and the Onsala Space Observatory} Submillimeter Heterodyne instrument
\citep[FLASH;][]{2006A&A...454L..21H} on APEX
\citep{2006A&A...454L..13G}. The two lines separated by about 40 MHz
were covered by one of the MPIfR Fast Fourier Transform Spectrometers
(XFTTS) connected to the 0.8 mm receiver. The original spectral
resolution at 333 GHz is about 35 m\,s$^{-1}$; the spectra shown
in this paper are Hanning smoothed to a resolution of 69
m\,s$^{-1}$ (76 kHz). 
The APEX beam size (FWHM) is $\sim$20$\arcsec$ at 333 GHz.  The
observations were carried out between 29 and 31 May, 2015 in stable
and fairly good weather conditions (PWV 0.7-1.2 mm), using position
switching for sky subtraction. The average system temperature at 333
GHz was 260 K.  The resulting RMS noise at a resolution of 69
m\,s$^{-1}$ is 0.017 K on the $T_{\rm A}^*$ scale.

\subsection{IRAM 30m}

The $1_{11}$-$1_{01}$ lines of o/p-NH$_2$D
at $\sim$86 and $\sim$110 GHz were observed at the IRAM 30m
telescope using the EMIR 090
receiver\footnote{www.iram.es/IRAMES/mainWiki/EmirforAstronomers} and
the VESPA autocorrelator.
The spectral resolution of this instrument, 20 kHz, is 
68 m\,s$^{-1}$ at 86 GHz and $53$ m\,s$^{-1}$ at 110 GHz.  At
these frequencies, the beam sizes of the telescope are $29\arcsec$ and
$23\arcsec$, respectively.
The observations were performed on July 5, 2015, in acceptable weather
conditions (PWV 8--10~mm). The observing mode was position
switching. The integration times for the o/p-NH$_2$D lines
were 15 and 22 minutes, and the average system temperatures at 86
and 110 GHz were 170 K and 250 K, respectively.  The resulting RMS
noise levels of the o/p--NH$_2$D spectra were 0.07 K and
0.08 K on the $T_{\rm A}^*$ scale.

\section{Hyperfine pattern calculations}\label{sec:spectro}

Just like in NH$_3$, the nitrogen nucleus in NH$_2$D can tunnel
across the plane made by the H and D nuclei.  Each rotational
transition is thus split into a doublet by this inversion
motion. The resulting rotation-inversion levels
are either symmetric or anti-symmetric under the exchange
of the two protons and they will either correspond
to para or ortho level.

Hyperfine patterns were calculated from a fit of high-resolution
data that can be divided into three sets. 
The first set consists of microwave and far infrared transitions involving the 
two inversion substates of the ground vibrational state.
This first set includes 174 microwave transitions \citep{cohen_pickett_1982}
and 297 far infrared transitions \citep{fusina_88} for
which no hyperfine structure is resolved.  The second set
involves the 76 microwave transitions listed in Table~IX of
\cite{cohen_pickett_1982} for which the nitrogen atom quadrupole
coupling structure is resolved.  The last data set comprises
the 21 hyperfine components measured by \cite{kukolich1968}
for the para $4_{14}$-$4_{04}$ rotation-inversion
transition at 25\,023.8~MHz. For this last data set, the
quadrupole coupling structure due to both the nitrogen and
deuterium atoms is resolved.

Rotation-inversion energies were computed with the help of a
semi-rigid rotator Hamiltonian for both inversion substates and
a second order Coriolis coupling term between these substates
\citep{cohen_pickett_1982}.  Molecule-fixed components of
the hyperfine coupling tensors are written using the IAM axis
system of this reference. Quadrupole and magnetic spin-rotation
hyperfine couplings were taken into account leading to a
hyperfine Hamiltonian depending on four coupling constants:
two for the quadrupole coupling of the nitrogen and deuterium
atoms and two for the magnetic spin-rotation coupling of the
same atoms.  Using equations similar to Eqs.~(3) and (17)
of \cite{thaddeus_krisher_loubser}, these four constants can
be expressed as diagonal matrix elements of four operators
involving four rank two hyperfine coupling tensors: the
zero trace ${\bf\chi}^{{\rm N}}$ and ${\bf\chi}^{{\rm D}}$,
describing quadrupole coupling of the nitrogen and deuterium
atoms, respectively, and ${\bf C}^{{\rm N}}$ and ${\bf C}^{{\rm
D}}$, corresponding to the magnetic spin-rotation coupling of
the same atoms.

Hyperfine energies were calculated taking the coupling scheme:
$ {\bf J} + {\bf I}_{{\rm N}} = {\bf F}_1$ and ${\bf F}_1 +
{\bf I}_{{\rm D}} = {\bf F}$ where ${\bf J}$ is the rotational
angular momentum, and ${\bf I}_{{\rm N}}$ and ${\bf I}_{{\rm
D}}$ are angular momenta of the nitrogen and deuterium nuclei,
respectively. The corresponding coupled basis set functions
$|J,I_{{\rm N}}; F_1, I_{{\rm D}}; F, M_F\rangle$ were used
to setup the hyperfine Hamiltonian matrix.  Matrix elements
were taken from \cite{thaddeus_krisher_loubser}.

The inversion-rotation transitions belonging to the first
data set were analyzed first allowing us to determine a
set of spectroscopic parameters analogous to that listed
in Table~III of \cite{cohen_pickett_1982}.  Components of
the four hyperfine coupling tensors were then fitted to the
frequencies of transitions belonging to the second and third
data sets evaluating the hyperfine coupling constants with the
eigenfunctions retrieved in the first analysis.  For lines
belonging to the first (second) data set, an experimental
uncertainty value of 0.1~MHz (1~kHz) was assumed and compares
well with a root mean square deviation of the observed minus
calculated residual of 8.4~kHz (1.5~kHz).  Table~\ref{eqq_value}
reports the values obtained for fitted components of the
hyperfine coupling tensors.  These components are given
in the axis system of \cite{cohen_pickett_1982}. Due
to the fact that the rotation-inversion transition measured
by \cite{kukolich1968} is characterized by $\Delta J = 0$,
calculated hyperfine frequencies mainly depend on the sum
$\chi^{{\rm D}}_{xx} + \chi^{{\rm D}}_{yy}$. For this reason,
only $\chi^{{\rm D}}_{yy}$ was varied in the analysis and
$\chi^{{\rm D}}_{xx}$ was constrained to a value retrieved
from the deuteron coupling constant $(eq_{\xi}Q)_D$ reported
by \cite{kukolich1968} and using the fact that angle between
the ND bond and the $z$-axis \citep{cohen_pickett_1982}
is 78.98$^{\circ}$. Magnetic spin-rotation coupling effects
could only be retrieved for the nitrogen atom. As all three
diagonal components of the corresponding
tensor could not be determined separately,
they were constrained to be equal, as for the normal species \citep{kukolich_1967}.

The results of the above analysis were used to predict hyperfine
patterns of other rotation-inversion transitions,
evaluating hyperfine intensities with Eq.~(29) of
\cite{thaddeus_krisher_loubser}. We note that hyperfine effects
due to the two hydrogen atoms may be important for ortho
transitions. These effects were evaluated taking into account
the spin-spin coupling, calculated from the equilibrium structure
of the molecule, and the spin-rotation coupling, evaluating the
coupling constant for $J$=1 from \cite{kukolich_1967}
and \cite{garvey_1976}.  Inclusion of these additional hyperfine
couplings were found to affect marginally the line parameters and
in particular, the linewidth is at most altered by a few 
percents (see Section~4). Therefore, these effects have been neglected in the following.

\begin{table}
\caption{\label{eqq_value}Hyperfine parameters.}
\centering
\begin{tabular}{l c l c}
\hline\hline
Parameter & Value & Parameter & Value \\ \hline
$\chi^{{\rm N}}_{xx}/ {\rm MHz} $ & 1.906(84) & $\chi^{{\rm D}}_{xx}/ {\rm kHz} $ & 274.67\tablefootmark{a} \\
$\chi^{{\rm N}}_{yy}/ {\rm MHz} $ & 2.040(98) & $\chi^{{\rm D}}_{yy}/ {\rm kHz}$ & \minnumb114.9(15) \\
$C^{{\rm N}}_{xx} / {\rm kHz}$ & 4.993(100)\tablefootmark{b} &                    &              \\
\hline
\end{tabular}
\tablefoot{Numbers in parentheses
are one standard deviation in the same units as the last
digit. \tablefoottext{a}{Constrained value.}
\tablefoottext{b}{The three diagonal components
of this tensor were constrained
to be equal.}}
\end{table}

\section{Modelling} \label{sec:models}

The HFS method of the CLASS
software\footnote{the HFS acronym stands for "HyperFine Structure" and
  a description of the method is available at
  www.iram.es/IRAMES/otherDocuments/postscripts/classHFS.ps}
allows to quickly analyse hyperfine spectra.  In
particular, it gives some basic parameters
of the lines as the width $\Delta v$ of the individual
transitions or the opacity summed over all the hyperfine
components $\tau_{tot}$.
The total column density of a molecule can be inferred from the parameters obtained with the HFS fit using relation \citep[see e.g.][]{bacmann2010,mangum2015}
\begin{eqnarray} \label{eq1}
N = \frac{8 \pi \nu^3}{c^3} \frac{Q(T_{ex})}{g_{u} \, A_{ul}} 
\frac{exp\left(\frac{E_u}{k_b T_{ex}}\right)}
{exp\left(\frac{h\nu}{k_b T_{ex}}\right)-1}
\int \tau_{ul} \, \textrm{d}v.
\end{eqnarray}
As explained in \citet{bacmann2010}, the integration of the opacity
over velocity for the $u \to l$ component can be related to the total
opacity $\tau_{tot}$ given by the HFS fit, through
\begin{eqnarray} \label{eq2}
\int \tau_{ul} \, \textrm{d}v \propto \tau_{tot} \,  s_{ul} \, \Delta v,
\end{eqnarray}
where $s_{ul}$ is the line-strength associated with the isolated
hyperfine transition.  The linewidth which enters this expression is
associated to thermal and non-thermal processes, i.e.  the motion of
molecules at microscopic (temperature) and macroscopic (turbulence)
scales.  Such an expression is routinely used in astrophysical
applications. In the particular case of NH$_2$D, such a relation is
often applied to the analysis of the 86 or 110 GHz lines, with
the aim to put constraints on the deuterium fraction
\citep[see e.g.][]{olberg1985,tine2000,roueff2005,busquet2010,fontani2015}.

In the case of NH$_2$D, the hyperfine structure induced by the D nucleus
is not resolved in astrophysical media since the broadening of
the lines due to non-thermal motions is larger than the
hyperfine splitting.  Hence, 
it seems reasonable to analyze the lines just taking into account the
hyperfine structure induced by the $^{14}$N nucleus. Doing so, the HFS
method applied to the H-MM1 observations described in the previous
section lead to derive the following parameters for the p-NH$_2$D
lines :
\begin{itemize}
\item 110 GHz : $\tau_{tot}$ =  2.2$\pm 0.6$ and $\Delta v$ = 0.33$\pm 0.02$ km s$^{-1}$
\item 333 GHz : $\tau_{tot}$ =  2.0$\pm 0.8$  and $\Delta v$ = 0.20$\pm 0.02$ km s$^{-1}$
\end{itemize}
and for the o-NH$_2$D lines :
\begin{itemize}
\item 86 GHz : $\tau_{tot}$ =  5.1$\pm 0.3$ and $\Delta v$ = 0.37$\pm 0.01$ km s$^{-1}$
\item 333 GHz : $\tau_{tot}$ =  2.2$\pm 0.3$ and $\Delta v$ = 0.24$\pm 0.01$ km s$^{-1}$
\end{itemize}
Different transitions of a molecule should have
similar intrinsic linewidths if they originate from the same region of
the cloud. In astrophysical sources, this condition is not necessarily
fulfilled: if the source harbours density or temperature gradients,
lines with different critical densities are formed in different parts
of the cloud.  The factor $\sim$1.6$\pm0.3$ found between the linewidths of
the $1_{11}$-$1_{01}$ and $1_{01}$-$0_{00}$ transitions of 
o/p-NH$_2$D is, however, puzzling because all four transitions have
high critical densities ($7\,10^4$--$7\,10^5$ cm$^{-3}$), all the
observations have similar spatial resolutions (see Sect. \ref{sec:observations}), 
and finally, because NH$_2$D should be strongly concentrated on the centre of the core for
chemical reasons. We would thus expect these lines to probe the same volume of gas and we should, in principle,
 derive similar values for $\Delta v$.

\begin{figure}
\begin{center}
\includegraphics[angle=0,scale=.27]{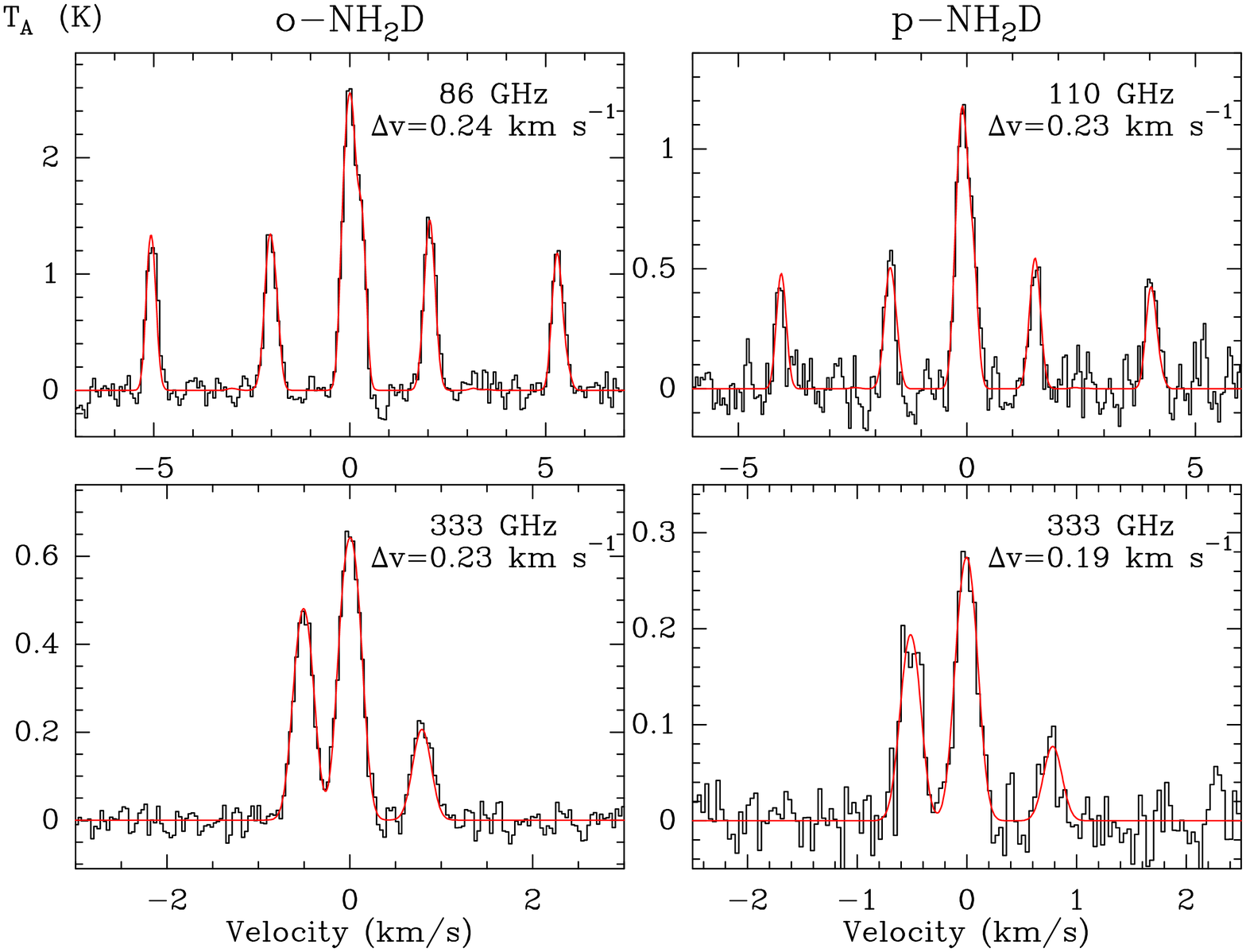} 
\caption{The left column shows the fit of the o-NH$_2$D lines at 86 GHz (upper panel) and 333 GHz (lower panel) with the hyperfine structure due to the D nucleus. The right column shows the fit of the p-NH$_2$D lines at 110 GHz (upper panel) and 333 GHz (lower panel). Note that fits of similar quality are obtained if the hyperfine structure due to D is omitted.} \vspace{-0.1cm}
\label{fig:pNH2D_HFS}
\end{center}
\end{figure}

\begin{figure}
\begin{center}
\includegraphics[angle=0,scale=.31]{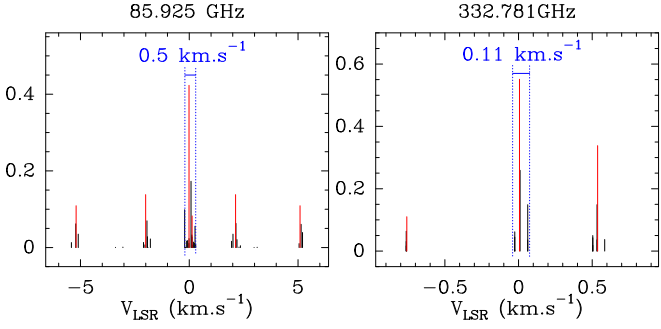} 
\caption{Line-strengths, normalized and centered on rest frequencies, for the 86 GHz and 333 GHz lines of o--NH$_2$D. 
In black, the spectroscopy accounts for the coupling with D while the spectroscopy
obtained just accounting for N is shown in red.} \vspace{-0.1cm}
\label{fig:oNH2D_spectro}
\end{center}
\end{figure}

By taking into account the splitting induced by the D nucleus (see
Table~\ref{table2} and \ref{table3}), the parameters derived from the HFS method are, for
p--NH$_2$D:
\begin{itemize}
\item 110 GHz : $\tau_{tot}$ =  2.2$\pm$0.7 and $\Delta v$ =  0.23$\pm0.02$ km s$^{-1}$
\item 333 GHz : $\tau_{tot}$ =  1.4$\pm$0.8 and $\Delta v$ = 0.19$\pm0.02$ km s$^{-1}$
\end{itemize}
and for the o-NH$_2$D :
\begin{itemize}
\item 86 GHz   : $\tau_{tot}$ =  5.2$\pm$0.5 and $\Delta v$ =  0.24$\pm0.01$ km s$^{-1}$
\item 333 GHz : $\tau_{tot}$ =  2.3$\pm$0.4 and $\Delta v$ =  0.23$\pm0.01$ km s$^{-1}$
\end{itemize}
The corresponding fits of the o/p-NH$_2$D hyperfine transitions are
compared to the observations in Fig. \ref{fig:pNH2D_HFS}.  For both
spin isomers, we find that the differences between the linewidths are
largely reduced, the differences being now at most $\sim$20\%. In
particular, we find that the effect is most important for the
86 and 110 GHz lines while the two lines at 333 GHz have similar
linewidths with or without taking into account the hyperfine structure
induced by the D nucleus.  This difference between the lines
comes from the spectroscopy and is illustrated in
Fig. \ref{fig:oNH2D_spectro} for the o-NH$_2$D spin isomer (the
description that follows would be similar for p-NH$_2$D). 
In this figure, we see that the number of hyperfine components is limited for
the $1_{01}$-$0_{00}$ transitions.
 Additionally, for these lines, the spread in velocity of the
hyperfine components is reduced by comparison to the spread of
hyperfines in the $1_{11}$-$1_{01}$ transitions. 
Hence, for the
latter transitions, taking into account the coupling with D leads to
a reduced $\Delta v$. Finally, we see that for the 86 and 110 GHz
transitions, the estimates of $\tau_{tot}$ are similar with the two
sets of line lists. As a result, according to Eq. \ref{eq1} and
\ref{eq2}, the error made in the linewidth estimate will translate
directly to the column density estimate. In the case of the H-MM1
observations, the two estimates will typically differ by a factor of
$\sim$1.5, which is well above calibration
errors.

\section{Conclusion} \label{sec:conclusion}
 
 We have observed the $1_{11}$-$1_{01}$ and $1_{01}$-$0_{00}$ lines of
o/p-NH$_2$D towards the prestellar core H-MM1, and
calculated the hyperfine patterns of the observed transitions. We
found that when the hyperfine splitting induced by the D nucleus is
neglected (as done in previous studies), the line analysis leads to
inconsistent results pertaining the linewidths of the two transitions
of o/p-NH$_2$D. For both spin isomers, the widths of the
$1_{11}$-$1_{01}$ and $1_{01}$-$0_{00}$ lines differ by a factor of
1.6$\pm$0.3 if the D coupling is not taken into account. On the contrary,
 the new line list gives
comparable linewidths for all the transitions. An error in
the linewidth will be transferred to the column density estimate, and
the effect is particularly pronounced for the $1_{11}$-$1_{01}$ lines of
ortho and para-NH$_2$D at 86 and 110 GHz. In the case of H-MM1, the
column density estimates derived from these lines with or without the
hyperfine structure owing to D differ by a factor of $\sim$1.5.

\begin{table}[htdp]
\scriptsize
\begin{center}
\begin{tabular}{cc|cc||cl||cl}
\multicolumn{2}{c}{1$_{11}$} & \multicolumn{2}{c}{1$_{01}$} & \multicolumn{2}{c}{p-NH$_2$D} 
& \multicolumn{2}{c}{o--NH$_2$D} \vspace{0.05cm} \\ \hline
F$_1$ & F & F$_1$' & F'  & $\nu$ (MHz) & A$_{ul}$ (s$^{-1}$) & $\nu$ (MHz) & A$_{ul}$ (s$^{-1}$)  \\ \hline
  0 &  1 &  1 &  0 & 110151.982 &      1.95 ( 6) & 85924.691 &      9.23 ( 7) \\
 0 &  1 &  1 &  2 & 110152.040 &      9.18 ( 6) & 85924.749 &      4.35 ( 6) \\
 0 &  1 &  1 &  1 & 110152.072 &      5.20 ( 6) & 85924.781 &      2.46 ( 6) \\
 0 &  1 &  2 &  2 & 110152.565 &      6.21 ( 8) & 85925.273 &      2.93 ( 8) \\
 0 &  1 &  2 &  1 & 110152.662 &      1.52 ( 7) & 85925.370 &      7.17 ( 8) \\
 2 &  1 &  1 &  0 & 110152.935 &      1.99 ( 6) & 85925.644 &      9.43 ( 7) \\
 2 &  2 &  1 &  2 & 110152.954 &      6.05 ( 7) & 85925.662 &      2.87 ( 7) \\
 2 &  3 &  1 &  2 & 110152.980 &      4.44 ( 6) & 85925.688 &      2.10 ( 6) \\
 2 &  2 &  1 &  1 & 110152.986 &      2.52 ( 6) & 85925.694 &      1.20 ( 6) \\
 2 &  1 &  1 &  2 & 110152.993 &      4.85 ( 8) & 85925.702 &      2.30 ( 8) \\
 2 &  1 &  1 &  1 & 110153.025 &      3.36 ( 6) & 85925.734 &      1.59 ( 6) \\
 2 &  2 &  2 &  2 & 110153.478 &      8.81 ( 6) & 85926.186 &      4.18 ( 6) \\
 0 &  1 &  0 &  1 & 110153.484 &      2.82 (10) & 85926.191 &      1.33 (10) \\
 2 &  3 &  2 &  2 & 110153.504 &      1.07 ( 6) & 85926.212 &      5.09 ( 7) \\
 2 &  1 &  2 &  2 & 110153.517 &      2.72 ( 6) & 85926.225 &      1.29 ( 6) \\
 1 &  1 &  1 &  0 & 110153.534 &      1.58 ( 6) & 85926.243 &      7.47 ( 7) \\
 2 &  2 &  2 &  3 & 110153.536 &      2.09 ( 6) & 85926.244 &      9.92 ( 7) \\
 2 &  3 &  2 &  3 & 110153.562 &      1.10 ( 5) & 85926.270 &      5.23 ( 6) \\
 1 &  2 &  1 &  2 & 110153.574 &      2.92 ( 6) & 85926.282 &      1.38 ( 6) \\
 2 &  2 &  2 &  1 & 110153.576 &      2.48 ( 6) & 85926.284 &      1.18 ( 6) \\
 1 &  0 &  1 &  1 & 110153.580 &      2.73 ( 6) & 85926.288 &      1.30 ( 6) \\
 1 &  1 &  1 &  2 & 110153.592 &      2.10 ( 6) & 85926.301 &      9.96 ( 7) \\
 1 &  2 &  1 &  1 & 110153.606 &      1.04 ( 6) & 85926.314 &      4.91 ( 7) \\
 2 &  1 &  2 &  1 & 110153.615 &      8.30 ( 6) & 85926.323 &      3.94 ( 6) \\
 1 &  1 &  1 &  1 & 110153.625 &      1.15 ( 6) & 85926.333 &      5.42 ( 7) \\
 1 &  2 &  2 &  2 & 110154.098 &      1.45 ( 6) & 85926.806 &      6.87 ( 7) \\
 1 &  1 &  2 &  2 & 110154.117 &      5.18 ( 6) & 85926.825 &      2.46 ( 6) \\
 1 &  2 &  2 &  3 & 110154.156 &      5.63 ( 6) & 85926.864 &      2.67 ( 6) \\
 1 &  0 &  2 &  1 & 110154.170 &      9.22 ( 6) & 85926.877 &      4.37 ( 6) \\
 1 &  2 &  2 &  1 & 110154.196 &      1.11 ( 7) & 85926.904 &      5.25 ( 8) \\
 1 &  1 &  2 &  1 & 110154.215 &      6.91 ( 7) & 85926.922 &      3.28 ( 7) \\
 2 &  2 &  0 &  1 & 110154.397 &      2.64 ( 8) & 85927.104 &      1.25 ( 8) \\
 2 &  1 &  0 &  1 & 110154.437 &      1.24 ( 7) & 85927.143 &      5.85 ( 8) \\
 1 &  0 &  0 &  1 & 110154.991 &      4.59 ( 6) & 85927.698 &      2.18 ( 6) \\
 1 &  2 &  0 &  1 & 110155.017 &      5.40 ( 6) & 85927.724 &      2.56 ( 6) \\
 1 &  1 &  0 &  1 & 110155.036 &      5.85 ( 6) & 85927.743 &      2.77 ( 6) \\
\end{tabular}
\end{center}
\caption{Line list of the 1$_{11}$-1$_{01}$ o/p-NH$_2$D transitions, with the hyperfine structure due to both the
nitrogen and deuterium nuclei. The A$_{ul}$ coefficients are given in the format $a(b)$ such that A$_{ul}$ = $a \, 10^{-b}$.}
\label{table2}
\end{table}%

\begin{table}[htdp]
\scriptsize
\begin{center}
\begin{tabular}{cc|cc||cl||cl}
\multicolumn{2}{c}{1$_{01}$} & \multicolumn{2}{c}{0$_{00}$} & \multicolumn{2}{c}{o-NH$_2$D} 
& \multicolumn{2}{c}{p--NH$_2$D} \vspace{0.05cm} \\ \hline
F$_1$ & F & F$_1$' & F'  & $\nu$ (MHz) & A$_{ul}$ (s$^{-1}$) & $\nu$ (MHz) & A$_{ul}$ (s$^{-1}$)  \\ \hline
 0 &  1 &  1 &  2 & 332780.875 &      4.68 ( 6) &332821.618 &      4.37 ( 6) \\
 0 &  1 &  1 &  1 & 332780.875 &      2.26 ( 6) &332821.618 &      2.11 ( 6) \\
 0 &  1 &  1 &  0 & 332780.875 &      1.20 ( 6) &332821.618 &      1.12 ( 6) \\
 2 &  1 &  1 &  2 & 332781.695 &      3.14 ( 7) &332822.439 &      2.93 ( 7) \\
 2 &  1 &  1 &  1 & 332781.695 &      4.54 ( 6) &332822.439 &      4.24 ( 6) \\
 2 &  1 &  1 &  0 & 332781.695 &      3.29 ( 6) &332822.439 &      3.07 ( 6) \\
 2 &  3 &  1 &  2 & 332781.735 &      8.14 ( 6) &332822.479 &      7.60 ( 6) \\
 2 &  2 &  1 &  2 & 332781.793 &      1.59 ( 6) &332822.537 &      1.48 ( 6) \\
 2 &  2 &  1 &  1 & 332781.793 &      6.56 ( 6) &332822.537 &      6.12 ( 6) \\
 1 &  1 &  1 &  1 & 332782.285 &      1.35 ( 6) &332823.029 &      1.25 ( 6) \\
 1 &  1 &  1 &  2 & 332782.285 &      3.15 ( 6) &332823.029 &      2.94 ( 6) \\
 1 &  1 &  1 &  0 & 332782.285 &      3.65 ( 6) &332823.029 &      3.41 ( 6) \\
 1 &  2 &  1 &  1 & 332782.317 &      1.59 ( 6) &332823.062 &      1.48 ( 6) \\
 1 &  2 &  1 &  2 & 332782.317 &      6.56 ( 6) &332823.062 &      6.12 ( 6) \\
 1 &  0 &  1 &  1 & 332782.375 &      8.14 ( 6) &332823.120 &      7.60 ( 6) \\
\end{tabular}
\end{center}
\caption{Same as Table~\ref{table2} but for the 1$_{01}$-0$_{00}$ o/p-NH$_2$D transitions.}
\label{table3}
\end{table}%

\begin{acknowledgements}
This work has been supported by the Agence Nationale de la Recherche
(ANR-HYDRIDES), contract ANR-12-BS05-0011-01 and by the CNRS national program
``Physico-Chimie du Milieu Interstellaire''.
AP, PC and JP acknowledge the financial support of the European Research Council (ERC; project PALs 320620).
JH acknowledges support from the MPE and the Academy of Finland
grant 258769. Partial salary support for AP was provided by a Canadian Institute for Theoretical Astrophysics (CITA) National Fellowship.
\end{acknowledgements}

\bibliographystyle{aa}
\bibliography{biblitex}

\begin{thebibliography}{22}
\expandafter\ifx\csname natexlab\endcsname\relax\def\natexlab#1{#1}\fi

\bibitem[{{Bacmann} {et~al.}(2010){Bacmann}, {Caux}, {Hily-Blant}, {Parise},
  {Pagani}, {Bottinelli}, {Maret}, {Vastel}, {Ceccarelli}, {Cernicharo},
  {Henning}, {Castets}, {Coutens}, {Bergin}, {Blake}, {Crimier}, {Demyk},
  {Dominik}, {Gerin}, {Hennebelle}, {Kahane}, {Klotz}, {Melnick}, {Schilke},
  {Wakelam}, {Walters}, {Baudry}, {Bell}, {Benedettini}, {Boogert}, {Cabrit},
  {Caselli}, {Codella}, {Comito}, {Encrenaz}, {Falgarone}, {Fuente},
  {Goldsmith}, {Helmich}, {Herbst}, {Jacq}, {Kama}, {Langer}, {Lefloch}, {Lis},
  {Lord}, {Lorenzani}, {Neufeld}, {Nisini}, {Pacheco}, {Pearson}, {Phillips},
  {Salez}, {Saraceno}, {Schuster}, {Tielens}, {van der Tak}, {van der Wiel},
  {Viti}, {Wyrowski}, {Yorke}, {Faure}, {Benz}, {Coeur-Joly}, {Cros},
  {G{\"u}sten}, \& {Ravera}}]{bacmann2010}
{Bacmann}, A., {Caux}, E., {Hily-Blant}, P., {et~al.} 2010, \aap, 521, L42

\bibitem[{{Busquet} {et~al.}(2010){Busquet}, {Palau}, {Estalella}, {Girart},
  {S{\'a}nchez-Monge}, {Viti}, {Ho}, \& {Zhang}}]{busquet2010}
{Busquet}, G., {Palau}, A., {Estalella}, R., {et~al.} 2010, \aap, 517, L6

\bibitem[{{Caselli} \& {Dore}(2005)}]{caselli2005}
{Caselli}, P. \& {Dore}, L. 2005, \aap, 433, 1145

\bibitem[{Cohen \& Pickett(1982)}]{cohen_pickett_1982}
Cohen, E.~A. \& Pickett, H.~M. 1982, J.\ Mol.\ Spectrosc., 93, 83

\bibitem[{{Cohen} \& {Pickett}(1982)}]{cohen1982}
{Cohen}, E.~A. \& {Pickett}, H.~M. 1982, Journal of Molecular Spectroscopy, 93,
  83

\bibitem[{{Coudert} \& {Roueff}(2006)}]{coudert2006}
{Coudert}, L.~H. \& {Roueff}, E. 2006, \aap, 449, 855

\bibitem[{{Coudert} \& {Roueff}(2009)}]{coudert2009}
{Coudert}, L.~H. \& {Roueff}, E. 2009, \aap, 499, 347

\bibitem[{{De Lucia} \& {Helminger}(1975)}]{delucia1975}
{De Lucia}, F.~C. \& {Helminger}, P. 1975, Journal of Molecular Spectroscopy,
  54, 200

\bibitem[{{Fontani} {et~al.}(2015){Fontani}, {Busquet}, {Palau}, {Caselli},
  {S{\'a}nchez-Monge}, {Tan}, \& {Audard}}]{fontani2015}
{Fontani}, F., {Busquet}, G., {Palau}, A., {et~al.} 2015, \aap, 575, A87

\bibitem[{Fusina {et~al.}(1988)Fusina, {Di Lonardo}, Johns, \&
  Halonen}]{fusina_88}
Fusina, L., {Di Lonardo}, G., Johns, J. W.~C., \& Halonen, L. 1988, J.\ Mol.\
  Spectrosc., 127, 240

\bibitem[{Garvey {et~al.}(1976)Garvey, Lucia, \& Cederberg}]{garvey_1976}
Garvey, R.~M., Lucia, F. C.~D., \& Cederberg, J.~W. 1976, Molec.\ Phys., 31,
  265

\bibitem[{{G{\"u}sten} {et~al.}(2006){G{\"u}sten}, {Nyman}, {Schilke},
  {Menten}, {Cesarsky}, \& {Booth}}]{2006A&A...454L..13G}
{G{\"u}sten}, R., {Nyman}, L.~{\AA}., {Schilke}, P., {et~al.} 2006, \aap, 454,
  L13

\bibitem[{{Heyminck} {et~al.}(2006){Heyminck}, {Kasemann}, {G{\"u}sten}, {de
  Lange}, \& {Graf}}]{2006A&A...454L..21H}
{Heyminck}, S., {Kasemann}, C., {G{\"u}sten}, R., {de Lange}, G., \& {Graf},
  U.~U. 2006, \aap, 454, L21

\bibitem[{{Johnstone} {et~al.}(2004){Johnstone}, {Di Francesco}, \&
  {Kirk}}]{2004ApJ...611L..45J}
{Johnstone}, D., {Di Francesco}, J., \& {Kirk}, H. 2004, \apjl, 611, L45

\bibitem[{Kukolich(1967)}]{kukolich_1967}
Kukolich, S.~G. 1967, Phys.\ Rev., 156, 83

\bibitem[{{Kukolich}(1968)}]{kukolich1968}
{Kukolich}, S.~G. 1968, \jcp, 49, 5523

\bibitem[{{Mangum} \& {Shirley}(2015)}]{mangum2015}
{Mangum}, J.~G. \& {Shirley}, Y.~L. 2015, \pasp, 127, 266

\bibitem[{{Olberg} {et~al.}(1985){Olberg}, {Bester}, {Rau}, {Pauls},
  {Winnewisser}, {Johansson}, \& {Hjalmarson}}]{olberg1985}
{Olberg}, M., {Bester}, M., {Rau}, G., {et~al.} 1985, \aap, 142, L1

\bibitem[{{Parise} {et~al.}(2011){Parise}, {Belloche}, {Du}, {G{\"u}sten}, \&
  {Menten}}]{2011A&A...528C...2P}
{Parise}, B., {Belloche}, A., {Du}, F., {G{\"u}sten}, R., \& {Menten}, K.~M.
  2011, \aap, 528, C2

\bibitem[{{Roueff} {et~al.}(2005){Roueff}, {Lis}, {van der Tak}, {Gerin}, \&
  {Goldsmith}}]{roueff2005}
{Roueff}, E., {Lis}, D.~C., {van der Tak}, F.~F.~S., {Gerin}, M., \&
  {Goldsmith}, P.~F. 2005, \aap, 438, 585

\bibitem[{Thaddeus {et~al.}(1964)Thaddeus, Krisher, \&
  Loubser}]{thaddeus_krisher_loubser}
Thaddeus, P., Krisher, L.~C., \& Loubser, J. H.~N. 1964, J.\ Chem.\ Phys., 40,
  257

\bibitem[{{Tin{\'e}} {et~al.}(2000){Tin{\'e}}, {Roueff}, {Falgarone}, {Gerin},
  \& {Pineau des For{\^e}ts}}]{tine2000}
{Tin{\'e}}, S., {Roueff}, E., {Falgarone}, E., {Gerin}, M., \& {Pineau des
  For{\^e}ts}, G. 2000, \aap, 356, 1039

\end{thebibliography}

\end{document}